# Accurate estimation of measurement position in Brillouin optical correlation-domain reflectometry based on Rayleigh noise spectral analysis

Keita Kikuchi, Ryo Inoue, Haruki Sasage,
Heeyoung Lee, and Yosuke Mizuno

*Abstract*— Brillouin optical correlation-domain reflectometry (BOCDR) is unique in its ability to measure distributed strain and temperature changes along a fiber under test (FUT) from a single end, offering random access and relatively high spatial resolution, making it promising for infrastructure monitoring. BOCDR achieves spatial resolution through frequency modulation of the laser output, and this modulation frequency determines the measurement position, necessitating accurate association of modulation frequencies with positions on the FUT. However, a practical method to precisely correlate modulation frequency values with FUT positions has not yet been proposed. This study introduces a method leveraging the change in Rayleigh noise spectrum with modulation frequency to accurately associate these frequencies with positions on the FUT. The effectiveness of this method is proved through distributed strain measurement.

## I. Introduction

Distributed strain and temperature sensors utilizing Brillouin scattering in optical fibers, which operate based on the frequency shift of the scattered spectrum and are unaffected by optical power variations, have been promoted for infrastructure monitoring [1], [2], [3], [4], [5]. Particularly, distributed measurement techniques based on the correlation-domain method have attracted attention for their ability to achieve both relatively high spatial resolution and random-access capabilities, compared to the time-domain [6], [7], [8], [9] and frequency-domain methods [10], [11]. To date, two types of measurement techniques using the correlation-domain method have been proposed: analysis and reflectometry [12], [13], [14], [15], [16], [17], [18], [19], [20], [21], [22], [23], [24], [25], [26]. Brillouin optical correlation-domain analysis (BOCDA) introduces frequency-modulated pump and probe lights from both ends of the fiber under test (FUT), inducing stimulated Brillouin scattering at a specific location through their interaction [14], [15], [16], [17], [18], [19]. In contrast, Brillouin optical correlation-domain reflectometry (BOCDR) operates by injecting modulated light from only one end of the FUT, selectively extracting spontaneous-Brillouin-scattered light from a specific location [12], [13], [14], [20], [21], [22], [23], [24], [25], [26]. Although BOCDA offers higher signal-to-noise ratios (SNRs) and superior measurement performance, as BOCDR is more advantageous in terms of implementation cost and installation convenience, this paper focuses on BOCDR.

In BOCDR, frequency-modulated light is split into two paths: one serves as the reference and the other as the incident light [12], [13]. The incident light is introduced into the FUT via a circulator, and the Brillouin-scattered light returning from the FUT interferes with the reference light. A so-called correlation peak [12], [13], [14], which serves as a measurement point, is created at a location where there is no phase difference in the frequency modulation due to the difference in the two light paths, i.e., where the correlation between the two lights is highest. When the frequency modulation is periodic, multiple correlation peaks are generated at equal intervals along the FUT. By limiting the length of the FUT, it is possible to generate only one correlation peak within the FUT, allowing for the selective detection of Brillouin-scattered light at that position. Furthermore, by controlling the modulation frequency, it is possible to sweep the correlation peak along the FUT, enabling the distributed measurement of the Brillouin gain spectrum (BGS) and the Brillouin frequency shift (BFS) across the entire length of the FUT.

As the position of the correlation peak (measurement point) on the FUT is determined by the modulation frequency, for accurate measurement of the BGS and BFS distributions along the FUT, it is essential to precisely match the modulation frequency value with the sensing position. Conventionally, the presence of a correlation peak at the FUT's end face was assumed based on the modulation frequency when the BGS was observed at a 3 dB down power level. However, this method, based on optical power, suffers from low positional accuracy. Thus, there is a demand for a concrete method to accurately match the position of correlation peak on the FUT with the modulation frequency.

In this work, we propose a method that utilizes the variation of the Rayleigh scattering-induced noise spectrum with the modulation frequency value in BOCDR to accurately correlate the position of a correlation peak on the FUT with the modulation frequency. We also demonstrate the effectiveness

This work was partly supported by the JSPS KAKENHI Grant Nos. 21H04555 and 22K14272 (Corresponding author: Y. Mizuno).

K. Kikuchi, R. Inoue, H. Sasage, and H. Lee are with the Graduate School of Engineering and Science, Shibaura Institute of Technology, Tokyo 135-8548, Japan (e-mail: af20098@shibaura-it.ac.jp; af20077@shibaura-it.ac.jp; ma22068@shibaura-it.ac.jp; hylee@shibaura-it.ac.jp).

Y. Mizuno is with the Faculty of Engineering, Yokohama National University, Yokohama 240-8501, Japan (e-mail: mizuno-yosuke-rg@ynu.ac.jp). He is also with the Institute for Multidisciplinary Sciences, Yokohama National University.

of this method by distributed strain measurement. Building on preliminary findings from our earlier papers [25,26], we advance a refined technique for exact measurement point estimation within the FUT, enhancing BOCDR's accuracy and utility in distributed sensing.

## II. PRINCIPLES

Figure 1 illustrates the concept of standard BOCDR based on self-heterodyne detection [12], [13]. A laser outputs light frequency-modulated by a sinusoidal wave, which is divided into pump and reference lights. The pump light is injected into the FUT via a circulator. The Brillouin-scattered light (Stokes light) returning from the FUT interferes with the reference light, and the optical beat signal is converted into an electrical signal using a photodetector (PD). This process allows for selective extraction and observation of the BGS at a point where the frequency difference between the two lights does not vary over time, i.e., at a correlation peak. The peak frequency of this BGS represents the BFS, which linearly depends on the strain and temperature changes applied to the FUT. By controlling the modulation frequency to sweep the correlation peak (sensing point) along the FUT, the distributions of the BGS and the BFS can be measured, enabling the measurement of distributed strain and temperature changes along the FUT.

In BOCDR, the measurement range $d_m$ (distance between adjacent correlation peaks) and the spatial resolution $\Delta z$ are given by [12], [13]

$$d_m = \frac{c}{2nf_m}, \tag{1}$$

$$\Delta z = \frac{c\Delta\nu_B}{2\pi n f_m \Delta f}, \tag{2}$$

where the parameters involved include the speed of light ($c$), Brillouin linewidth ($\Delta\nu_B$), core refractive index ($n$), modulation frequency ($f_m$), and modulation amplitude ($\Delta f$). These equations underscore the trade-off relationship between achieving a broad measurement range and maintaining high spatial resolution in BOCDR.

Figure 2 illustrates the spread of the Rayleigh scattering noise spectrum and the BGS observed by an ESA [12], [13], [25]. The spread of the BGS results from the interference of frequency-modulated Brillouin-scattered light, generated at various positions within the FUT, with the reference light. Similarly, the spread of the noise spectrum from 0 Hz occurs due to the interference of frequency-modulated Rayleigh-scattered light (at each position within the FUT) with the reference light [12], [13], [25]. The spectral width of the Rayleigh noise, $W_R$, is determined by the situation where the phase difference in frequency modulation between the Rayleigh-scattered light and the reference light becomes maximal. Therefore, $W_R$ is given by [13]

$$W_R = 2\Delta f \max\left\{\sin\left(\frac{\varphi}{2}\right)\right\}, \tag{3}$$

where $\varphi$ represents the phase difference in frequency modulation between the Rayleigh-scattered light and the reference light, ranging from 0 to $\pi$ (correlation peaks

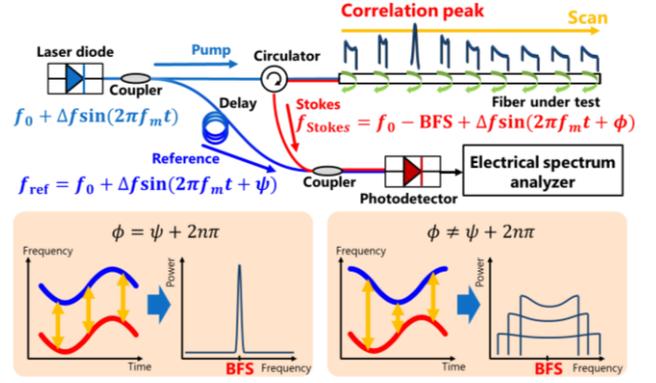
Fig. 1. Conceptual illustration of BOCDR. BFS: Brillouin frequency shift.

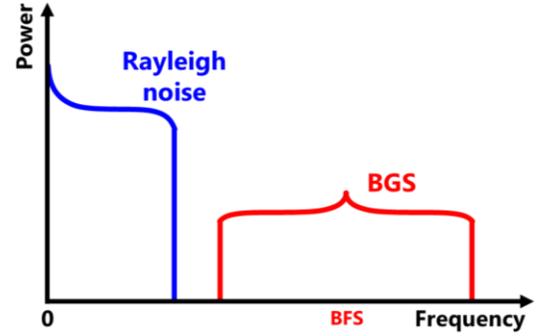
Fig. 2. Schematic electrical spectrum of the BOCDR output.

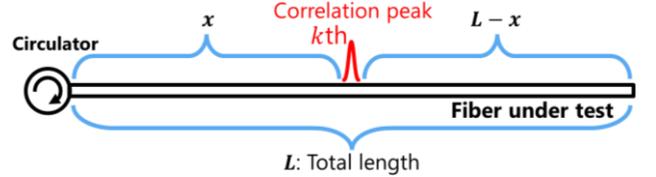
Fig. 3. Definitions of $L$ and $x$ in Eqs. (4)–(8).

correspond to locations where $\varphi = 0$). If the FUT includes a position where $\varphi = \pi$, the Rayleigh noise spectral width becomes twice the modulation amplitude, indicating that the modulation amplitude must be set to less than half the value of the BFS when such a condition is met within the FUT. However, when the FUT length is shorter than half the measurement range, and thus does not include a position where $\varphi = \pi$, it is possible to set the modulation amplitude to greater than half the value of the BFS, improving spatial resolution [13]. This discussion focuses on cases where the FUT length is shorter than half the measurement range.

When the FUT length is shorter than half the measurement range, the modulation amplitude value estimated from the Rayleigh noise spectrum width is reported as follows (refer to Fig. 3) [25,26]:

$$\Delta f_R = \frac{W_R}{2} \cdot \operatorname{cosec}\left\{\frac{\pi}{d_m} \max(L-x, x)\right\}, \tag{4}$$

where $L$ is the length of the FUT, and $x$ is the position along the FUT (i.e., the distance from the proximal end of the FUT). Figure 4 schematically shows the relationship between the observed Rayleigh noise spectrum in BOCDR and the position of the correlation peak on the FUT. As depicted, the presence of a correlation peak in the FUT and the resultant spectra at two locations lead to a two-tier structure in the Rayleigh noise spectrum. The second tier's spectrum width is given by:

$$W_{\text{R\_2nd}} = \frac{\Delta f}{2} \cdot \sin\left\{\frac{\pi}{d_m} \max(L-x, x)\right\}, \quad (5)$$

and the first tier's component by:

$$W_{\text{R\_1st}} = \frac{\Delta f}{2} \cdot \sin\left\{\frac{\pi}{d_m} \min(L-x, x)\right\}. \quad (6)$$

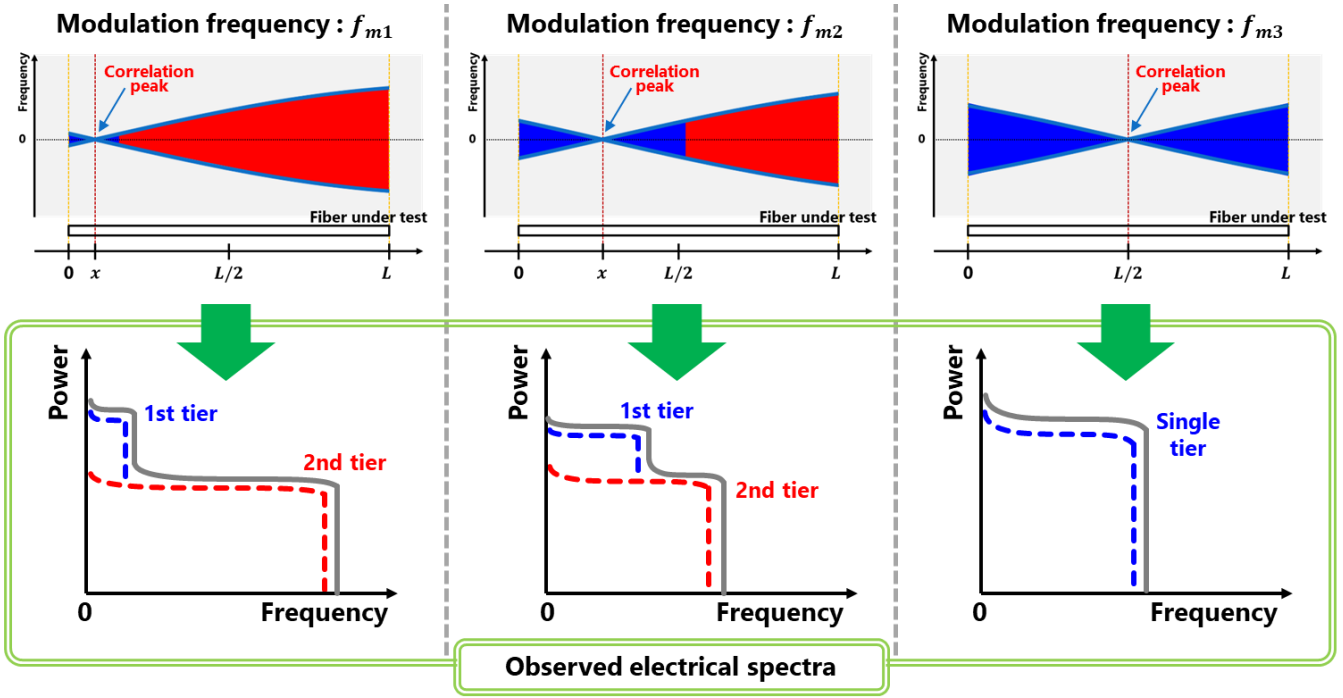

Fig. 4. Relationship between correlation peak positions and Rayleigh noise spectra in BOCDR. The upper figures show the beat spectra.

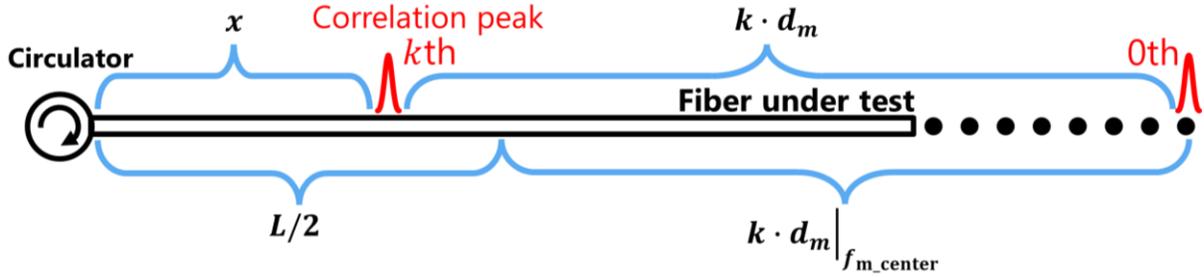

Fig. 5. Measurement position calculated from the Rayleigh noise spectrum.

If the correlation peak is located at the center of the FUT, the Rayleigh noise spectrum becomes a single-tier structure with a width of

$$W_{\text{R\_single}} = \frac{\Delta f}{2} \cdot \sin\left\{\frac{\pi}{d_m} \cdot \frac{L}{2}\right\}. \quad (7)$$

Therefore, when the Rayleigh noise spectrum changes from a two-tier to a single-tier structure, it indicates that the correlation peak is at the FUT's center. Thus, the modulation frequency value $f_{\text{m\_center}}$ at this moment allows for the calculation of any position $x$ along the FUT as follows (see Fig. 5):

$$x = k \cdot d_m\big|_{f_{\text{m\_center}}} + L/2 - k \cdot d_m, \quad (8)$$

where $k$ is the order of the correlation peak.

### III. EXPERIMENTS

#### A. Experimental setup

Figure 6 shows the BOCDR setup used in experiments, which employed a distributed-feedback laser with 1550 nm. The pump light, Stokes light, and reference light were each amplified to 24.0 dBm, 3.8 dBm, and 3.1 dBm, respectively, using EDFAs. A polarization scrambler was employed to minimize the signal fluctuations due to polarization variations. A delay line of approximately 120 m was inserted in the reference path to ensure that only one -1st-order correlation peak existed within the FUT. The configuration of the FUT is shown in Fig. 7, with a total length of 21 m, including the second port of the circulator. A bending loss was applied near the distal end of the FUT to prevent Fresnel reflection. In addition, a strain of 0.15% was applied to a 1.0 m section around the center to demonstrate distributed BFS measurement.

#### B. Spectral analysis

First, to calculate the modulation frequency when the correlation peak is located at the center of the FUT, the observation of the Rayleigh noise spectrum was conducted. The resolution bandwidth (RBW), video bandwidth (VBW), sweep time, and averaging count of the ESA were set to 3 MHz, 1 kHz, 150 msec, and 40 times, respectively. The peak-to-peak voltage of the function generator (FG) was fixed at 1.0 $V_{pp}$, and the change from a two-tier to a one-tier Rayleigh noise spectrum was observed under three conditions with different step widths for the FG's modulation frequency: (a) 1.9–2.0 MHz (every 0.1 MHz), (b) 2.06–2.08 MHz (every 0.01 MHz), and (c) 2.076–2.078 MHz (every 0.001 MHz).

The observed Rayleigh noise spectra under different conditions are shown in Figs. 8(a)–(c). As seen in the waveform in Fig. 8(a), a third tier appears in the Rayleigh noise spectrum. This tier, caused by the interference between weak reflected light from the circulator's first to third port and Rayleigh-scattered light in the FUT [27], is excluded from the measurement target in this method, which is based on the changes corresponding to the correlation peak used for distribution measurement. From the modulation frequency values obtained in (a), further observations of the Rayleigh noise spectrum were made with narrower steps in (b) and (c), resulting in a modulation frequency of 2.078 MHz where the spectrum changes from two-tier to one-tier. Considering the frequency step of 0.001 MHz in (c), the error is estimated to be a maximum of ±~3 cm for this case.

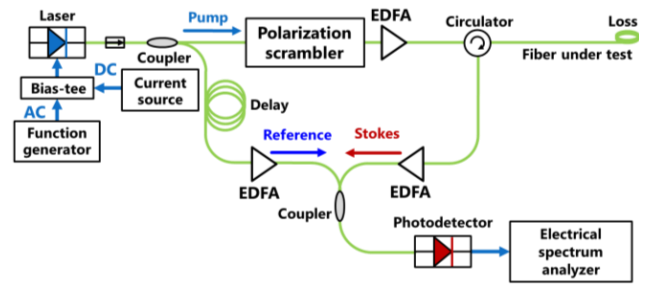

Fig. 6. Experimental setup of BOCDR. AC: alternating current, DC: direct current, EDFA: erbium-doped fiber amplifier, PSCR: polarization scrambler.

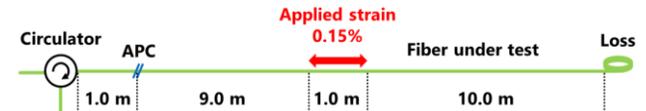

Fig. 7. Structure of the FUT. APC: angled physical contact.

### C. Distributed stain measurement

To verify this approach, distributed strain measurement was conducted under conditions derived from the aforementioned experiment. Using the modulation frequency value of 2.078 MHz, determined to position the correlation peak at the FUT's center, and Eq. (9), the correlation peak was swept across the 9–12 m section of the FUT. The FG's voltage was set at 1.4 $V_{pp}$. At the FUT center, the modulation amplitude was 2.136 GHz, resulting in a spatial resolution of approximately 22 cm and a measurement range of about 40 m. The ESA's RBW, VBW, sweep time, and the average count were set to 3 MHz, 1 kHz, 30 msec, and 60 times, respectively, to capture the BGS at each modulation frequency within the sweep range.

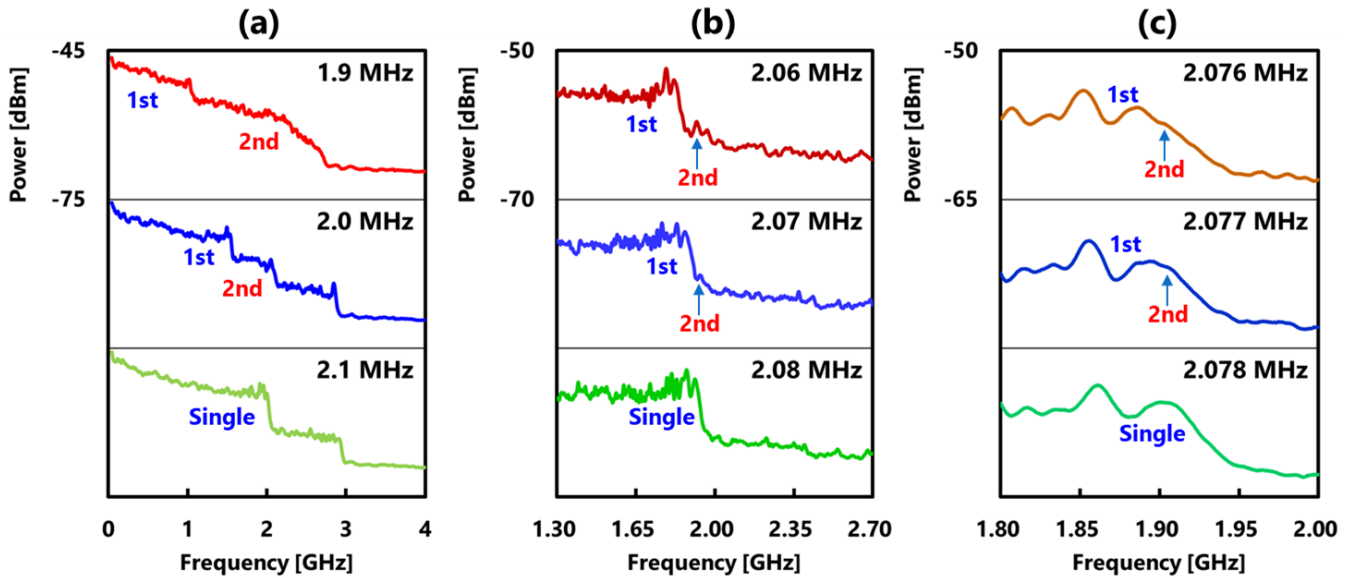

Fig. 8. Measured Rayleigh noise spectra at different modulation frequencies of (a) 1.9–2.1 MHz, (b) 2.06–2.08 MHz, and (c) 2.076–2.078 MHz.

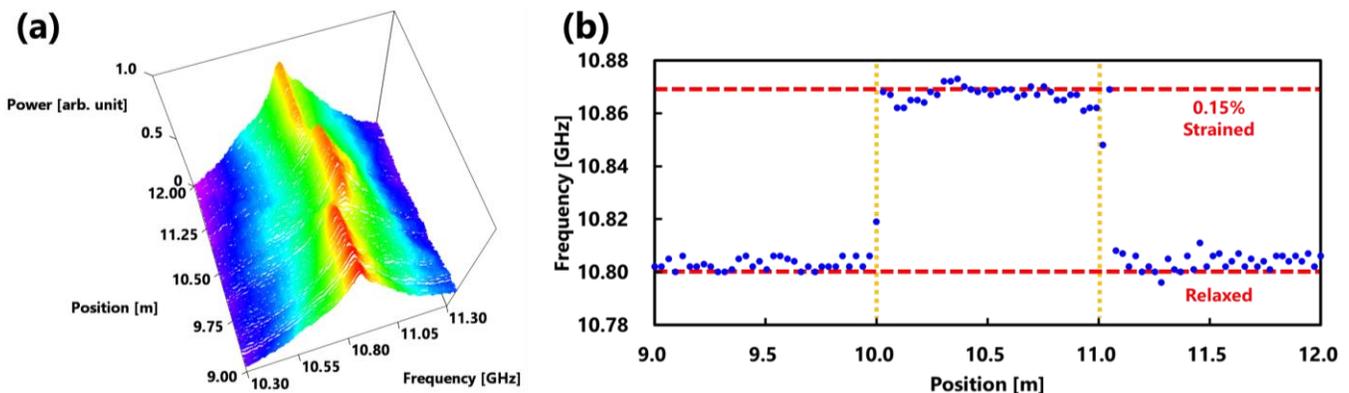

Fig. 9. Measured distributions of (a) BGS and (b) BFS.

Figure 9(a) displays the obtained BGS distribution, and Fig. 9(b) shows the BFS distribution derived by simply taking the frequency at which the BGS power was highest. An increase of approximately 70 MHz in the BFS, corresponding to a 0.15% strain, was observed in the 10–11 m section, showing an excellent agreement with the strain-applied section in the FUT. This result demonstrates the validity of our method for associating the position of the correlation peak on the FUT with the modulation frequency based on the Rayleigh noise spectral width.

## IV. Conclusion

We presented a method to accurately calculate the position of the correlation peak on the FUT from the modulation frequency by utilizing the Rayleigh noise spectral correspondence with the measurement location in BOCDR. Through experiments, we precisely calculated the modulation frequency at which the correlation peak is centered on the FUT and validated the effectiveness of our method with actual strain distribution measurement. This approach, which can be executed without any alterations to the existing experimental setup or additional expenses, is anticipated to serve as a fundamental technique for accurate distributed measurements in BOCDR.